\documentclass{article}
\begin{document}

\title{On the angular momentum of the neutron star}
\author{{Zakir F. Seidov} \thanks{%
e-mail:zakirs@yosh.ac.il} \\
Shemakha Astrophysical Observatory, Shemakha 373243, Azerbaijan,\\
Research Institute, College of Judea and Samaria, Ariel 44837, Israel\\}
\maketitle

\begin{abstract}
This is a more or less exact English version of the short note
published some 33 years ago in Russian.
Still it is of some interest and not only
from the historic point of view. No attempt was made
to add new references or upgrade the text.\\
It is shown that the mass loss due to rotation-driven hydrodynamical instability
during the catastrophic collapse of the star is small. Neutron star is formed
with a large rotational kinetic energy and the spin-down takes place
in the following life of neutron star as a pulsar.
\end{abstract}

\section{}

The Crab pulsar NP0532 with the smallest pulsation period ~.033 sec,
if it is connected with 1054 year SN, is also one of the youngest pulsars,
neutron stars (n.s.), remnants of SN explosion. If the pulsation period NP0532 is the
rotation period [1] then n.s. with rotation period $T=0.033$ sec, mass
$M\approx M_{\small \bigodot}$ and radius $R\simeq 10^{6}$ cm has rotational momentum less
than one-tenth the solar momentum. Evidently  or n.s. lost his momentum at
birth or the momentum loss took place during n.s. life. In [2], it is
claimed that the large momentum and mass loss is possible during the
catastrophic hydrodynamical instability. We reconsider the problem and
show that momentum loss is small and mass loss is negligible [11].

\section{}

We assume that the mass ejection of the rotating and contracting star takes
place at the star equator. Then for the momentum loss we have expression
$$d\,L=\alpha^2\,R^2\,\omega \, d\,M,$$
 where $\omega$ is the rotational angular velocity, $R$ is the equatorial
 radius, and we assume that the mass element leaves the star at distance
 $\alpha\,R>R,$
 due to e.g. magnetic forces confining the matter. \\
 Moreover we assume that all the way during the contraction, the injection
 condition $$\omega^{2}=G\,M/R^3$$ holds, where $G$ is the gravitational
 constant. Also we write the angular momentum relative the rotation axis in the
 form $$L=k\,M\,R^2\,\omega,$$ where $k$ is structural parameter depending on
 density distribution inside the star, and get finally:
 $$ {M\over M_0} =\left( R \over R_0 \right )^{\beta},$$
$${L\over L_0}= \left( R \over R_0 \right )^{3\,\beta /2+1/2},$$
$${\omega\over \omega_0}=\left( R \over R_0 \right )^{\beta /2-3/2},$$
$$\beta={k\over 2\,\alpha^2- 3\,k}.$$
\section{}
The values of $k$ were calculated for spherical configurations in [3],
and for  configurations at the state of rotational instability in [4].
In all cases $k < 1$; for
instance, for $n=3$ polytrope case (which is likely the case  for star losing
his gravitational stability), at $\alpha \approx 1$ and taking into account the
small deviations from the spherical symmetry, we have
$k=0.038,\ \ \beta=0.02.$

More particularly for a star with $ M\approx 2 \,10^{33}$ g, and $L=4\, 10^{48}$
erg sec, the condition of rotational instability commences first at $R_0=2.1\,
10^7$ cm, $\omega = 1.2\, 10^2\,\mbox {sec}^{-1}$. After contracting to state of n.s. with
$R=1.2\, 10^6$ cm, we get (assuming $\alpha=1$):
$$\omega_{n.s.}=8.8\, 10^3\,\mbox {sec}^{-1},\ \ T_{n.s.}=7\, 10^{-4}\,\mbox{sec}.$$
\section{}
For n.s. formed from a star with mass $M_0$, $\omega_{n.s.}\propto \omega_0
\,R_0^{3/2}\propto M_0^{1/2}$, that is the angular velocity of "new-born" n.s.
increases with increasing $M_0$ and does not depends on $L_0$. Also
$M_{n.s.}\approx M_0$ because the mass loss is negligible in all cases. Taking
into account the magnetic forces ($\alpha > 1$) leads to even lesser values of the
 momentum and mass loss. \\
 We conclude that the rotational instability can not be an effective mechanism
 of the momentum and mass loss at n.s. birth, see also [5,6,10].\\
 Therefore the neutron star (pulsar) is formed with a strong rotation and
 the momentum loss (spin-down) takes place during the pulsar's life. Of course we
 do not take into account the possibility of other mechanisms of momentum and mass
 loss due to e.g. explosions or any other mechanisms [6,7] due to uncertainty
 about their efficiency.\\
 As it is known [7,8,9], the different mechanisms of pulsar spin-down lead to
 different dependences of $d\,T/d\,t$ on time $t$. If we assume for evaluation
 purposes that $d\,T/d\,t=c\cdot T$, then we get $c=10^{-10}\,\mbox {sec}^{-1}$,
  which is close to value observed at the present.\\[5mm]
\parbox[t]{7cm}{
 Shemakha Astrophysical
 observatory,\\ Jan 1970}
\hfill \parbox[t]{3cm}{Z.F. Seidov}

\section{References}

1. T. Gold 1968, 1968Nature..218..731G\\
2. O.Gusejnov, F. Kasumov 1969, 1969ATsir...539..2G\\
ATsir = Astronomicheskij Tsirkulyar, Moscow, AN SSSR\\
3. Z. Seidov 1969, 1969Afz...5..503S\\
4. R. James 1964, 1964ApJ...140..552J\\
5. L. Auer, N. Woolf 1965, 1965ApJ..142..182A\\
6. V. Porfiryev 1969, 1969AZh...46..560P\\
7. I. Shklovsky 1969, 1969Atsir..494..2S\\
8. T. Gold 1969, 1969Natur..221...25G\\
9. I. Gunn, J. Ostriker 1969, 1969Natur..221..454G\\
10. V. Safronov 1951, 1951AZh ..28..244S\\
11. Z. Seidov, 1970, 1970ATsir..567..3S\\published 01 Jun 1970

\end{document}